\documentclass[aps,prl,reprint,superscriptaddress]{revtex4-1}

\usepackage{graphicx}
\usepackage{dcolumn}
\usepackage{bm}

\begin{document}


\title{Laser opacity in underdense preplasma of solid targets due to quantum
electrodynamics effects}

\author{W.-M. Wang}
\affiliation{Beijing National Laboratory for Condensed Matter
Physics, Institute of Physics, CAS, Beijing 100190, China}
\affiliation{Beijing Advanced Innovation Center for Imaging
Technology, Department of Physics, Capital Normal University,
Beijing 100048, China}
\author{P. Gibbon}
\affiliation{Forschungzentrum J\"ulich GmbH, Institute for Advanced
Simulation, J\"ulich Supercomputing Centre, D-52425 J\"ulich,
Germany} \affiliation{Centre for Mathematical Plasma Astrophysics,
Katholieke Universiteit Leuven, 3000 Leuven, Belgium}
\author{Z.-M. Sheng}
\affiliation{SUPA, Department of Physics, University of Strathclyde,
Glasgow G4 0NG, United Kingdom}\affiliation{Key Laboratory for Laser
Plasmas (MoE) and Department of Physics and Astronomy, Shanghai Jiao
Tong University, Shanghai 200240, China} \affiliation{IFSA
Collaborative Innovation Center, Shanghai Jiao Tong University,
Shanghai 200240, China}
\author{Y.-T. Li}
\affiliation{Beijing National Laboratory for Condensed Matter
Physics, Institute of Physics, CAS, Beijing 100190, China}
\affiliation{IFSA Collaborative Innovation Center, Shanghai Jiao
Tong University, Shanghai 200240, China} \affiliation{School of
Physical Sciences, University of Chinese Academy of Sciences,
Beijing 100190, China}
\author{J. Zhang}
\affiliation{Key Laboratory for Laser Plasmas (MoE) and Department
of Physics and Astronomy, Shanghai Jiao Tong University, Shanghai
200240, China} \affiliation{IFSA Collaborative Innovation Center,
Shanghai Jiao Tong University, Shanghai 200240, China}

\date{\today}

\begin{abstract}
We investigate how next-generation laser pulses at 10 PW $-$ 200 PW
interact with a solid target in the presence of a relativistically
underdense preplasma produced by amplified spontaneous emission
(ASE). Laser hole boring and relativistic transparency are strongly
restrained due to the generation of electron-positron pairs and
$\gamma$-ray photons via quantum electrodynamics (QED) processes. A
pair plasma with a density above the initial preplasma density is
formed, counteracting the electron-free channel produced by the hole
boring. This pair-dominated plasma can block the laser transport and
trigger an avalanche-like QED cascade, efficiently transfering the
laser energy to photons. This renders a 1-$\rm\mu m$-scalelength,
underdense preplasma completely opaque to laser pulses at this power
level. The QED-induced opacity therefore sets much higher contrast
requirements for such pulse in solid-target experiments than
expected by classical plasma physics. Our simulations show for
example, that proton acceleration from the rear of a solid with a
preplasma would be strongly impaired.
\end{abstract}


\maketitle

Current developments in ultraintense pulsed laser technology
indicate that laser pulses at the 100 PW level will be available in
the near future. A number of laboratories worldwide are pursuing
this high power frontier, such as the European Light Infrastructure
(ELI), which is designed to deliver laser pulses of 100-200 PW
\cite{ELI}, and the planned OMEGA EP-OPAL system which will supply
200 PW laser pulses \cite{OMEGA_EP_OPAL}. Interaction of such
ultra-relativistic pulses with matter will likely enter a
QED-dominant regime. The plasma electrons can be instantly
accelerated up to Lorentz factors $\gamma\sim a_0>1000$, where $a_0$
is the laser field strength normalized by $m_ec\omega_0/e$ and
$\omega_0$ is the laser frequency. The QED parameter
\cite{Erber,Elkina} of $\chi_e\simeq \gamma F_\perp/(eE_S)$ will
easily exceed 1, so that abundant $\gamma-$photons should be
generated via Compton scattering, potentially offering an
ultraintense $\gamma-$photon source \cite{Ridgers,Brady}. Here,
$E_S=1.32\times10^{18}V/m$ is the Schwinger field
\cite{Schwinger1,Schwinger2} and $F_\perp$ is the transverse
component of the Lorentz force. The generated photons of high energy
under the ultra-relativistic laser field, with the QED parameter of
photons \cite{Erber,Elkina} $\chi_{ph}\simeq (\hbar\omega/m_ec^2)
F_\perp/(eE_S)$ approaching or exceeding 1, will strongly trigger a
Breit-Wheeler process and create electron-positron pairs in an
avalanche-like way \cite{Bell,Fedotov,Bulanov,Nerush}, which can
present the laser intensity upper limit attainable
\cite{Fedotov,Bulanov} in the vacuum. This also provides a new
approach for pair source generation \cite{Sokolov,Ridgers,Lobet}.

On the other hand, when such a laser pulse is applied in a solid
target experiment, a preplasma produced by ASE could become more
unavoidable \cite{PM,ASE} than the case with a relatively low
intensity pulse available currently. For example, as low as
$10^{-11}$ ASE of a pulse at $10^{21}~\rm W cm^{-2}$ can only
produce a low level of preplasma ahead of a solid target
\cite{PM,ASE}, which can be ignored in most cases. For a pulse at
$10^{23}-10^{25}~\rm W cm^{-2}$, the same level of ASE (harder to
achieve in technology) can produce a significant level of preplasma
\cite{PM,ASE}. Due to the preplasma production, it will be a
challenge to make use of an extremely intense pulse in some key
applications based on laser interaction with solid-density targets,
such as ion acceleration via radiation pressure
\cite{Esirkepov04,Yan,Robinson,Shen,Yu,Zheng}, high-order harmonic
and attosecond pulse generation
\cite{Bulanov-HHG,Gibbon-HHG,Teubner}, or surface plasmon resonance
\cite{Kahaly,Wang,Monchoce}, ect. Therefore, it is important to
understand and anticipate the effects of a preplasma when a tightly
focussed pulse at 10$-$200 PW irradiates a solid target, where the
above-mentioned QED effects are expected to dominate.

In this Letter, we show that the QED effects can cause near complete
energy depletion of such a pulse in a relativistically-underdense,
small-scale preplasma. This contrasts to the regime studied to date
where the preplasma is actually rendered \textit{more} transparent
because of laser hole boring and relativistic self-induced
transparency \cite{Gibbon}. At an early stage in the interaction,
laser hole-boring acts within the leading edge of the pulse:
electrons are pushed away from the peak laser intensity zone by the
ponderomotive force and an electron-free channel is formed in the
preplasma. Under the combined charge-separation field and the
ponderomotive force, oscillating electrons leave the peak intensity
zone and cause abundant $\gamma-$photon generation, which cools the
electrons and reduces the relativistic transparency. Later, a large
number of pairs are created around the laser intensity peak zone,
which fill the preplasma ion channel. The pair plasma can have a
density much higher than the initial preplasma, which completely
hinders further laser hole boring. This pair plasma strongly absorbs
the laser energy and effectively transfers the energy to photons via
the avalanche-like QED cascade \cite{Bell,Fedotov,Bulanov,Nerush}.
In some case, the pair plasma density can be even higher than the
relativistic critical density \cite{Gibbon} and hence, the pulse is
strongly reflected, which could enhance the QED cascade with the
incident pulse together. The QED-induced inflation of the preplasma
implies that significant improvements in laser contrast technology
will be necessary before pulses with these intensities can be used
in solid-target experiments. In particular, only a small fraction of
the laser energy is finally absorbed by plasma electrons, which may
significantly limit the conventional applications based on electrons
pre-accelerated by laser pulses, e.g., ion acceleration.

Our investigation is carried out through particle-in-cell (PIC)
simulations with the two-dimensional (2D) version of the KLAPS code
\cite{KLAPS,KLAPS-QED}, in which $\gamma$-photon emission and pair
creation via QED effects are included. A laser pulse is incident
along the $+x$ direction with linear polarization along the y
direction, wavelength $1~\rm\mu m$ (or the period
$\tau_0=2\pi/\omega_0=3.33~\rm fs$), spot radius $r_0=1~\rm\mu m$,
and duration 30 fs of full width at half maximum. Its peak power
$P_0$ is 200 PW ($P_0$ from 10 PW to 180 PW will be also taken in
the simulations below) and peak amplitude $a_0=3049$. A 0.5-$\mu
m$-thick gold foil is taken, which is assumed to be composed of
$Au^{+10}$ ions and electrons with a density of 530 $n_c$
($n_c=1.1\times10^{21}\rm~cm^{-3}$) since we consider extremely
intense pulses (a case with $Au^{+15}$ ions and a higher electron
density will be discussed below). In front of the foil there is a
preplasma with an exponential density profile of a scalelength
$L=1~\rm\mu m$ ($L$ changing from $0.1~\rm\mu m$ to $0.9~\rm\mu m$
will also be taken below), which is expected to be produced by the
laser ASE. The simulation box size $82\rm\mu m \times 80\rm\mu m$ in
$x\times y$ directions is taken. Fourth-order algorithm for current
calculation is employed \cite{KLAPS,MA-FI-PRL}, with which the noise
is well controlled. We take a spatial resolution of 0.0208 $\rm\mu
m$ and 16 electrons and ions per cell. Because the particle number
is not large initially, memory overflow is avoided in our
simulations. The absorption budget of laser energy into the various
kinds of particles is closely monitored during the simulation.

\begin{figure}[htbp]
\includegraphics[width=3.4in]{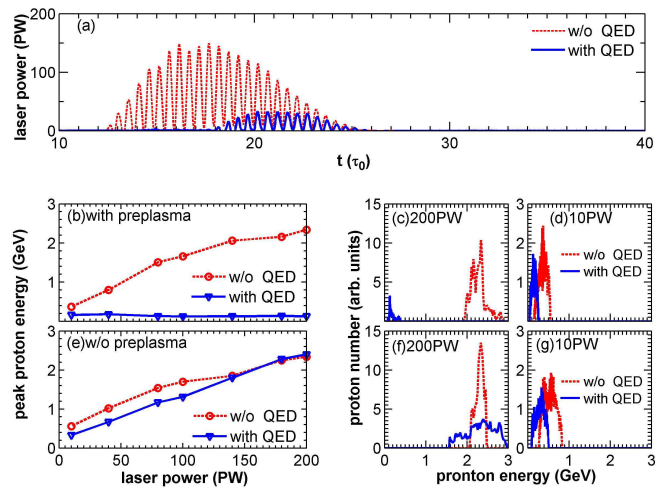}
\caption{\label{fig:epsart}(a) Temporal profile of the laser power
passing through the preplasma rear. In (a)-(f) the red broken line
and the blue solid line correspond to the simulations without and
with the QED effects, respectively. [(b),(e)] Peak energy of the
protons at 80 $\tau_0$ versus the laser power. [(c),(d),(f),(g)]
Proton energy spectra at 80 $\tau_0$ with different laser powers.
(b)-(d) are the results with a foil with a preplasma. (e)-(g) are
the results with a polished foil without a presplasma. }
\end{figure}

The red broken line in Fig. 1(a) is the simulation result without
the QED effects. It shows that due to laser hole boring and
relativistic transparency, the 200 PW laser pulse with $a_0=3049$
easily penetrates through the preplasma with an average density
$66n_c \ll a_0 n_c$ and a size of 8 $\rm \mu m$, retaining most of its
initial energy. However, the pulse loses nearly all energy in the same
preplasma if the QED effects are included, as seen in the blue line
in Fig. 1(a). Only 8\% initial laser energy (with 30 PW peak power)
is transported through the preplasma; without the QED effects, the
value is 65\% (with 150 PW peak power).

The large difference of laser energy depletion in the preplasma
results in quite different proton acceleration [see Figs. 1(b) and
1(c)], where a 0.05 $\mu m$ thick proton layer of 50 $n_c$ is
located in the rear of the gold foil and the layer is 2 $\mu m$ wide
in the y-direction. The protons are accelerated to a peak energy of
about 2.4 GeV at 80 $\tau_0$ without the QED effects (note that we
have not optimized target parameters for the proton acceleration).
With the QED effects, the peak energy is reduced to 0.13 GeV because
the depleted laser energy is mainly transferred to photons [see Fig.
2(a)], which do not contribute to the proton acceleration. As the
laser power is decreased [see Figs. 1(d) and 1(b)], the difference
of the proton energy is lessened between the two cases with and
without the QED effects, since the QED-induced depletion is
weakened. The protons are accelerated mainly via the target normal
sheath acceleration (TNSA) \cite{TNSA} since nearly all energy of
the pulse at 10-200 PW is depleted in the preplasma with the QED
effects (as discussed below). Without the QED effects, TNSA and
radiation pressure acceleration work together since much laser
energy is transported to the foil front.

In contrast to the results in Fig. 1(b) with the preplasma, Fig. 1(e) shows
that the peak energy is continuously enhanced
when a polished foil \textit{without} the preplasma is taken. In this case,
the peak energy shows little difference from the case without the QED
effects. This is because when the foil is thin enough, the reaction
of the foil to the pulse is weak and stays within a limited space, e.g.,
via the charge-separation field. In this case the foil electrons are mainly
accelerated along the laser propagation direction,
which is less effective in triggering Compton scattering for photon
generation (see $\chi_e\simeq \gamma F_\perp/(eE_S)$
\cite{Erber,Elkina}). Therefore, high energy protons/ions could be
generated by 100-PW-class pulses, provided the laser contrast is
sufficiently high and a thin enough foil is used. To optimize such
proton/ion acceleration, one could match the foil thickness and the
laser power as done in Ref. \cite{Esirkepov-scaling}, but only after taking QED
effects into account.

\begin{figure}[htbp]
\includegraphics[width=3.4in]{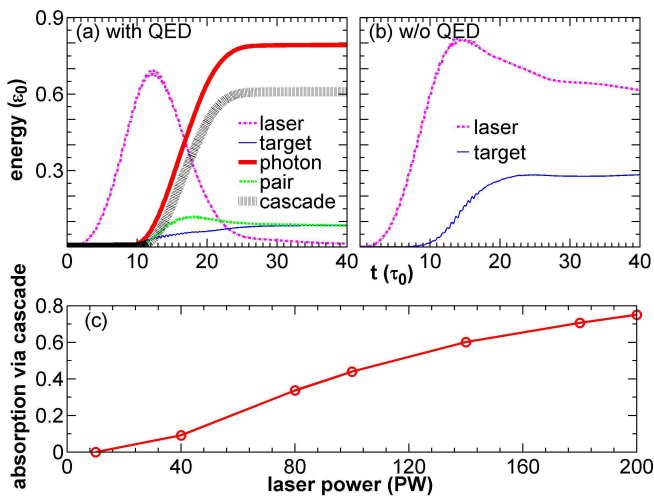}
\caption{\label{fig:epsart}[(a), (b)] Temporal evolution of residual
energies of the laser pulse, target particles, photons, pairs, and
the photons generated via the cascade (or only via the pairs),
normalized by the total laser energy $\varepsilon_0$, where the lase
power is taken as 200 PW. Plots (a) and (b) correspond to the
simulations with and without the QED effects, respectively. All
particles escaped away from the simulation box are recorded and
counted while the laser energy transported away is not counted.  (c)
Absorbed laser energy via the cascade in the preplasma region versus
laser powers, where the energy is normalized by the total laser
energy absorbed in this region.}
\end{figure}

In Fig. 2(a) the relative contributions of the generated photons and
pairs, and the target particles (mainly the preplasma) to the laser
energy depletion with the laser power of 200 PW are shown. About
80\% laser energy is finally transferred to $\gamma-$photons.
Firstly, some photons are generated around the laser axis at $y=0$
[see Fig. 3(i)] via Compton scattering of the preplasma electrons
after being accelerated by the pulse. The generated photons trigger
the Breit-Wheeler process and create pairs in the peak laser
intensity zone around the laser axis [see Fig. 3(i)]. Likewise, the
pairs under the laser fields also generate Compton photons. In this
way an avalanche-like cascade is formed causing copious photon
generation. According to the black broken line in Fig. 2(a), 77\% of
the photons are generated via the pairs (or via the cascade) and the
other 23\% via the preplasma electrons. Therefore, the pairs or the
cascade dominate the photon generation and the laser energy
absorption over the preplasma electrons.

\begin{figure}[htbp]
\includegraphics[width=3.4in]{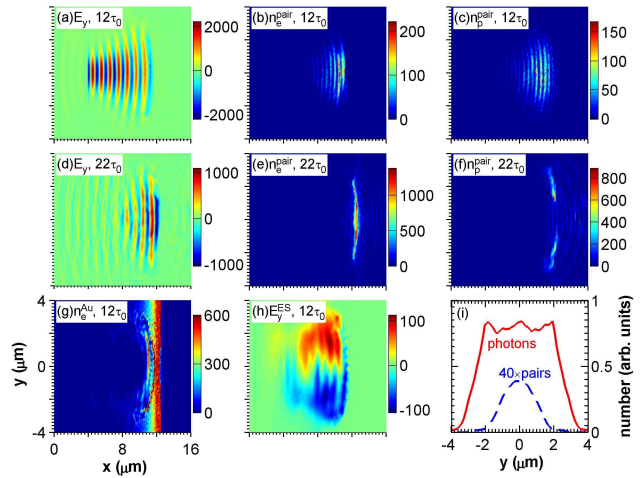}
\caption{\label{fig:epsart}Snapshots of laser electric fields
$eE_y/m_e \omega_0 c$ [(a), (d)], pair-plasma electron densities
$n_e^{pair}/n_c$ [(b), (e)] and positron densities $n_p^{pair}/n_c$
[(c), (f)] at different times, Au-target electron density
$n_e^{Au}/n_c$ (g), and electrostatic field $eE_y^{ES}/m_e \omega_0
c$ (h), where the maximum value is suppressed in (g) to show more
clearly. (i) Number of the generated photons and pairs versus the
transverse positions $y$, where the pair number is multiplied by a
factor of 40.}
\end{figure}

There are two reasons for the cascade dominating the laser
absorption. Firstly, the positrons, unlike massive ions, can be
easily accelerated to ultra-relativistic energies and can thus
contribute to the photon generation basically to the same extent as
the electrons; they can also inhibit the acceleration of ions and
protons. Secondly, due to requirement of the QED parameter
$\chi_{ph}\simeq (\hbar\omega/m_ec^2) F_\perp/(eE_S)$ approaching or
exceeding 1 \cite{Erber,Elkina} for strong creation of pairs, most
pairs are created in the peak laser intensity zone around the laser
axis [see Figs. 3(i), 3(b) and 3(c)], basically where the photon
generation rate is the highest. By contrast, the preplasma electrons
are expelled away from this peak intensity zone [see Fig. 3(g)] via
laser hole boring, within the pulse leading edge but mainly before
the pair creation. This causes the reduction of photon generation
via the preplasma electrons.

Due to the expelled electrons, a strong charge-separation field is
formed around the laser axis [see Fig. 3(h)], which tends to keep
the pair electrons within this region. Under this field, many
freshly created pair electrons remain localized, leading to a
growing pair-plasma electron density around the laser axis: it
reaches about 1400 $n_c$ at $22~\tau_0$, as seen in Figs. 3(e) and
3(b). Note that the charge-separation field also ensures that the
positron densities are always lower than the pair-plasma electron
densities, as seen in Figs. 3(c) and 3(f). In this way, the pair
plasma fills the preplasma-electron-free channel and its density is
much higher than the initial preplasma density, as seen in Figs.
3(g), 3(b), 3(c) and 3(e). Then, the laser hole boring and
relativistic transparency can be completely inhibited. As the pulse
strength $a_0$ is strongly reduced [see Fig. 3(d)], the density can
be even higher than $a_0n_c$ [see Fig. 3(e)], which causes the pulse
to be significantly reflected. The reflected pulse with the incident
one together can be quickly absorbed since they can strengthen the
QED cascade \cite{Bulanov,Nerush}.

The contribution of the QED cascade to the laser depletion is
weakened with decreasing laser power, as observed in Fig. 2(c). In
this figure, we plot the laser absorption via the cascade (or only
via the pairs). When the power is decreased to 100 PW from 200 PW,
the contribution of the pairs to the laser absorption is reduced to
44\% from 75\%. The value is further reduced to about 10\% at 40 PW.
Therefore, in a 1-$\rm\mu m$-scalelength preplasma, the cascade
starts to be important at 40 PW and becomes the leading effect at
around 100 PW.

\begin{figure}[htbp]
\includegraphics[width=3.4in]{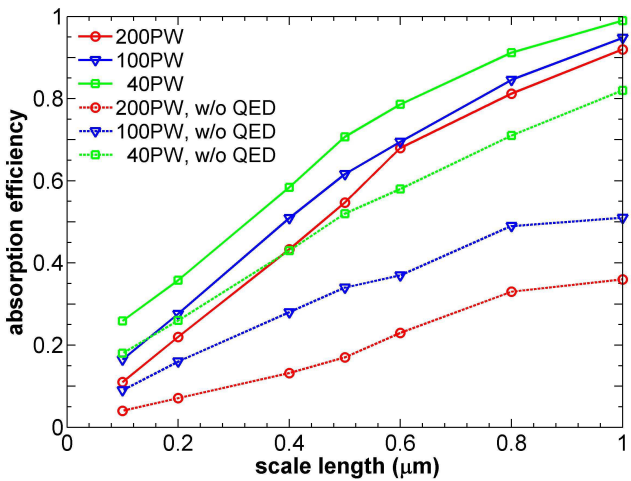}
\caption{\label{fig:epsart}The energy absorption efficiency of
lasers with different powers in the preplasma region versus
preplasma density scalelengths.}
\end{figure}

To gauge the sensitivity of this result to the initial preplasma
density scalelength $L$ and the laser power, a limited parameter
scan of absorption efficiencies is depicted in Fig. 4, about which
one can make several observations. First, one can see that QED
effects generally result in much higher laser absorption for powers
of 40-200 PW across the whole range of scalelengths considered ($L=
0.1-1.0 \rm\mu m$). Hence, much higher laser contrast will be
required to apply such a pulse in future experiments than expected
by classic plasma physics. The contrast between the two cases with
and without the QED effects is reduced for lower powers, but the
difference is still apparent even for relatively sharp profiles with
$L=~0.1 \rm\mu m$. Second, with a smaller $L$ (or a higher laser
contrast) and a given power, less laser energy is depleted in the
preplasma since the laser preplasma interaction zone is decreased.
Third, with a lower power and a given $L$, the absorption efficiency
of laser energy is higher. This is because the reaction of the
preplasma to the lower-power pulse is comparatively stronger, i.e.,
a higher ratio of charge-separation field strength to the pulse
strength and relatively more photons generated via the preplasma
electrons, even though fewer photons are generated via the pairs or
the cascade. To get 50\% of the laser energy transported to the
solid target front behind the preplasma according to Fig. 4, $L$
should be around $0.5\rm\mu m$ for the 200 PW pulse; and
$L\simeq0.4\rm\mu m$ for the 100 PW  and 40 PW pulses. When the
$Au^{+10}$ target is replaced by an $Au^{+15}$ target, the laser
depletion in the preplasma become stronger as seen in our
simulations. This is because the $Au^{+15}$ target and its preplasma
with the same $L$ have higher electron densities or more electrons.

\begin{figure}[htbp]
\includegraphics[width=3.4in]{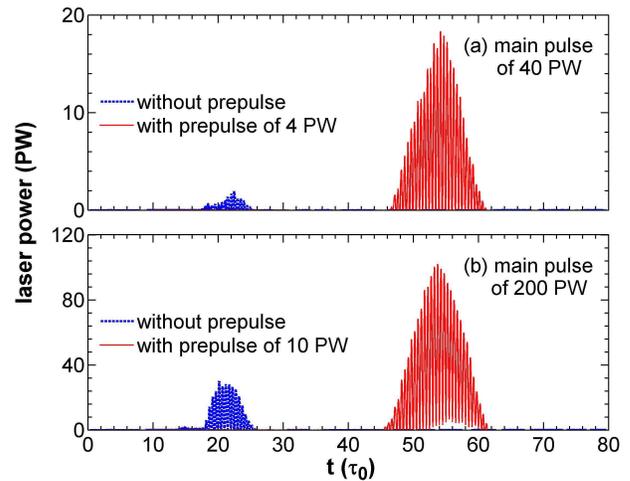}
\caption{\label{fig:epsart} Temporal profile of laser power passing
through the preplasma rear, where a controlled prepulse is taken 120
fs ahead of the main pulse in the red line and no prepulse is taken
in the blue broken line. (a) A 40 PW main pulse and a 4 PW prepulse
are adopted. (b) A 200 PW main pulse and a 10 PW prepulse are
adopted.}
\end{figure}

QED-induced preplasma inflation and the hinderance of the laser
passage to the solid target surface can be mitigated by using a
controlled prepulse. When the prepulse is ahead of the main one, it
can create a nearly electron-free channel in the preplasma, reducing
the interaction of the the main pulse with the preplasma electrons
and consequently weakening the QED effects. In our simulations, the
prepulse and main pulse have the same duration 30 fs and spot radius
$r_0=1~\rm\mu m$ and an $Au^{+10}$ target with a preplasma of
$L=1~\rm\mu m$ is employed. In Fig. 5(a), when a 4 PW prepulse is
taken, the 40 PW main pulse is transported through the preplasma
with 18 times more energy (up to 36\% initial energy of the main
pulse) than the case without a prepulse. In Fig. 5(b), when a 10 PW
prepulse is taken, the 200 PW main pulse is transported through the
preplasma with 6 times more energy (up to 47\% initial energy of the
main pulse) than the case without a prepulse.

In summary, we have shown by PIC simulation that the QED effects can
significantly enhance opacity of a laser pulse at 10-200 PW in a
relativistically-transparent preplasma produced by ASE nearly
unavoidable. A 1-$\rm\mu m$-scalelength preplasma is completely
opaque for such a pulse. To achieve transparencies above 50\%, a
preplasma with the scalelength below 0.4 $\rm\mu m$ is required.
Therefore, the QED-induced opacity sets much higher demands on laser
contrast technology for pulse powers of 10-200 PW than expected by
classic plasma physics. We have illustrated that ahead of the main
pulse, a controlled prepulse can effectively reduce such opacity,
even if the prepulse power is far below the main pulse power.

The QED-induced opacity is most potent when a high-density pair
plasma is formed around the peak laser intensity zone, to fill a
preplasma-electron-free channel produced via laser hole boring. The
pair plasma triggers an avalanche-like QED cascade to strongly
absorb the laser energy, which finally transfers to photons. The
cascade becomes the leading depletion mechanism and dominates over
the depletion by preplasma electrons when the pulse power reaches
100 PW. It starts to be significant at 40 PW. Below 40 PW, the laser
energy is depleted mainly due to pure Compton scattering via
preplasma electrons. In any case, little laser energy is finally
transferred to electrons of either the preplasma or the pairs, which
shall significantly affect the applications based on electrons
pre-accelerated by laser pulses, e.g., ion acceleration.

\begin{acknowledgments}
This work was supported by the National Basic Research Program of
China (Grants No. 2013CBA01500) and NSFC (Grants No. 11375261,
11421064, 11374210, and 113111048). ZMS acknowledges the support of
a Leverhulme Trust Research Grant and an EPSRC Grant No.
EP/N028694/1).
\end{acknowledgments}

\end{document}